\documentclass[aps,pra,preprintnumbers,showpacs,tightenlines]{revtex4}
\usepackage{amssymb}
\usepackage{amsmath}
\usepackage{graphicx}
\usepackage{epsfig}
\usepackage{subfigure}
\usepackage{amsfonts}
\usepackage{CJK}

\begin{document}

\title{Transferring multiqubit entanglement onto memory qubits in a decoherence-free subspace}

\author{Xiao-Ling He$^{1}$ and Chui-Ping Yang$^{2\star}$}

\address{$^1$School of Science, Zhejiang University of Science and Technology, Hangzhou, Zhejiang 310023, China}
\address{$^2$Department of Physics, Hangzhou Normal University, Hangzhou, Zhejiang 310036, China}
\address{$^\star$ yangcp@hznu.edu.cn}

\date{\today}

\begin{abstract}
Different from the previous works on generating entangled states, this work is focused on
how to transfer the prepared entangled states onto memory qubits for protecting them against decoherence. We here consider
a physical system consisting of $n$ operation qubits and $2n$ memory qubits placed in a cavity or coupled to a resonator.
A method is presented for transferring $n$-qubit Greenberger-Horne-Zeilinger (GHZ) entangled states
from the operation qubits (i.e., information processing cells) onto the memory qubits (i.e., information memory elements
with long decoherence time). The transferred GHZ states are encoded in a decoherence-free subspace against collective dephasing, and thus
can be immune from decoherence induced by a dephasing environment. In addition, the state transfer procedure has nothing to do with the
number of qubits, the operation time does not increase with the number of
qubits, and no measurement is needed for the state transfer. This proposal can be applied to a
wide range of hybrid qubits such as natural atoms and artificial atoms
(e.g., various solid-state qubits).
\end{abstract}

 \maketitle
\date{\today}

\section{Introduction}

Greenberger-Horne-Zeilinger (GHZ) states are of great interest in the
fundamental test of quantum mechanics [1] and may prove to be useful in quantum
metrology [2] and high-precision spectroscopy [3-5]. They have many
applications in quantum information processing (QIP) and quantum
communication such as quantum teleportation [6,7], entanglement swapping
[8], quantum cryptographic [9], and error correction protocols [10,11].
During the past decade, a great deal of efforts has been devoted to
generating GHZ states in various physical systems. For example, based on
cavity or circuit QED, many theoretical methods have been presented for
generating multiqubit GHZ states with atoms [12], quantum dots [13,14], and
superconducting qubits [15-18]. A natural question on how to store the
prepared GHZ states and how to protect them against decoherence would become
interesting and important.

Hybrid quantum systems, composed of different kinds of qubits, have
attracted tremendous attentions recently [19-27]. In addition, cavity or
circuit QED has been considered as one of the most promising candidates for
implementing large-scale QIP [28-30]. When two sets of qubits placed in a
cavity or resonator are hybrid (i.e., different types), qubits in one set
can act as information processing cells (i.e., the operation qubits) while
qubits in the other set play a role of information memory elements (i.e.,
the memory qubits). Here, the operation qubits are readily controlled and
thus used for performing quantum operations, while the memory qubits have
long decoherence time and thus are good to be used for storing quantum states.
It is well known that multiqubit entangled states are essential resources
for large-scale QIP. When performing QIP in a hybrid system, after a step
of information processing is completed, one needs to transfer entangled states
of operation qubits to memory qubits for storage; and one needs to transfer
the entangled states from the memory qubits back to
the operation qubits when a further step of processing is needed.

It has been recognized that decoherence resulting from the coupling of the
system with environment is one of the main obstacles in realizing a quantum
information processor. Generally speaking, the effect of decoherence in
carrying out QIP can be minimized as long as the typical operation time for
performing various quantum operations is much shorter than the decoherence
time of the system involved in QIP. In contrast, decoherence may pose a
significant issue during the storage of quantum states (especially in the
case when quantum states are stored for a long time). Hence, how to protect
quantum states from decoherence is important in achieving an efficient
storage of quantum states.

Motivated by the above, in this work we focus on how to transfer multiqubit
entangled states from operation qubits onto memory qubits, and how to
protect the transferred multiqubit entangled states against decoherence
induced due to interacting with phase-dumping environments during the
storage. Note that phase dumping can be a dominating decoherence source for
noise environments, which has been widely studied over the past years
[31-38].

In the following, we consider a physical system which consists of $n$
operation qubits and $2n$ memory qubits placed in a cavity or coupled to a
resonator. We will propose a method to transfer an $n$-qubit GHZ state $%
\alpha \left\vert 00...0\right\rangle +\beta \left\vert 11...1\right\rangle $
(with arbitrary coefficients $\alpha $ and $\beta $) from $n$ operation
qubits onto $2n$ memory qubits, in the form of $\alpha \left\vert
0101...01\right\rangle +\beta \left\vert 1010...10\right\rangle $ encoded in
a decoherence-free subspace (DFS) against collective dephasing. The DFS here is
spanned by the basis vectors $\left\vert 01\right\rangle $ and $\left\vert
10\right\rangle $ of every two memory qubits. As shown below, this approach
has the following advantages: (i) Since the transferred GHZ states
are encoded within a DFS, they are immune from dephasing during the storage on
the memory qubits; (ii) This protocol can be used to transfer the GHZ states
deterministically; (iii) The state transfer
procedure needs only a few basic operations, which does not depend on the
number of qubits; (iv) The operation time does not increase as the number of
qubits increases; (v) no measurement is needed; (vi) During the operation,
the level $\left\vert f\right\rangle $ only for two qubits is occupied and thus
decoherence from the qubits is greatly suppressed. This proposal
can be applied to implement the GHZ state transfer between
various hybrid qubits such as natural atoms and artificial atoms [e.g.,
quantum dots, nitrogen-vacancy (NV) centers, superconducting qubits, etc.].
To the best of our knowledge, how to transfer entangled states onto memory
qubits in a DFS, based on cavity/circuit QED, has not been investigated until today.

This paper is organized as follows. In Sect.~II, we show a general method to
transfer GHZ entangled states from $n$ operation qubits to
$2n$ memory qubits encoded in a DFS against collective
dephasing. In Sect.~III, we give a brief discussion on the experimental issues, we then
discuss the experimental feasibility of transferring a three-qubit GHZ
state from superconducting flux qubits to superconducting transmon qubits in
a DFS based on circuit QED. A concluding summary is presented in Sect.~IV.

\begin{figure}[tbp]
\begin{center}
\includegraphics[bb=153 534 481 669, width=8.5 cm, clip]{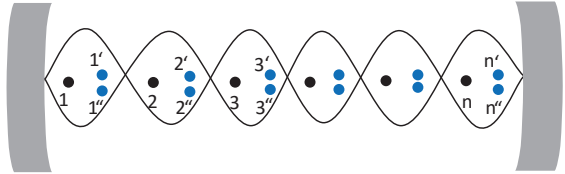} \vspace*{%
-0.08in}
\end{center}
\caption{(Color online) Diagram of operation qubits and memory qubits in a
cavity or a resonator. The two dashed curves represent the standing-wave
cavity mode. Each dark dot or blue dot represents a qubit. The dark dots
represent operation qubits ($1,2,...,n$) while the blue dots denote the
memory qubits ($1^{\prime },2^{\prime },...,n^{\prime },1^{\prime \prime
},2^{\prime \prime },...,n^{\prime \prime }$). Qubits \{$1^{\prime
},1^{\prime \prime }$\}, qubits \{$2^{\prime },2^{\prime \prime }$\},..., or
qubits \{$n^{\prime },n^{\prime \prime }$\} are arranged to be close
(compared to the environment's coherence length), such that the two memory
qubits in each pair couple to the environment in the same way appropriately.}
\label{fig:1}
\end{figure}

\section{Transferring GHZ states onto memory qubits in a DFS}

Consider two sets of hybrid qubits in a cavity or resonator (Fig. 1). The
first set contains $n$ identical operation qubits ($1,2,...,n$), the second set
contains $2n$ identical memory qubits ($1^{\prime },2^{\prime },...,n^{\prime },1^{\prime \prime },2^{\prime \prime
},...,n^{\prime \prime }$), but qubits in different sets are non-identical or hybrid.
Each qubit here is associated with a three-level quantum system (Fig. 2),
with the two levels $\left\vert g\right\rangle$ and $\left\vert
e\right\rangle $ representing a qubit and the third level $\left\vert
f\right\rangle $ used for the state manipulation. Before
the GHZ state transfer, the level structure of qubit $1$ is adjusted to be
different from that of qubits ($2,3,...,n$), and the level structure of qubit
$1^{\prime }$ is adjusted to be different from that of qubits ($2^{\prime
},3^{\prime },...,n^{\prime },1^{\prime \prime },2^{\prime \prime
},...,n^{\prime \prime }$) (Fig. 2). Note that the level spacings of
superconducting qubits can be rapidly (within a few nanoseconds [39-41)
adjusted by varying external control parameters (e.g., the magnetic flux
applied to the superconducting loop of qubits; see e.g. [28,39,42,43]), the
level spacings of NV centers can be adjusted by changing the
external magnetic field applied to the NV centers [44,45], and the level
spacings of atoms or quantum dots can be adjusted by changing the voltage on
the electrodes around each atom/quantum dot [46].

Each qubit is initially decoupled from the cavity (Fig.~2). Suppose that the
$n$ operation qubits ($1,2,...,n$) are initially in a GHZ state%
\begin{equation}
\left\vert \text{GHZ}\right\rangle _{12...n}=\alpha
\prod_{l=1}^{n}\left\vert g\right\rangle _{l}+\beta
\prod_{l=1}^{n}\left\vert e\right\rangle _{l}
\end{equation}
(with normalized unknown factors $\alpha $ and $\beta $)$.$

Note that the state $\left\vert \text{GHZ}\right\rangle _{12...n}$ of the $n$
operation qubits (1,2,...,$n$) can be prepared using the existing schemes (e.g.,
[12-18]). By applying classical pulses to the operation qubit $1,$ the states
$\left\vert g\right\rangle _{1}$ and $\left\vert e\right\rangle _{1}$ can be
easily converted into the states $\left\vert e\right\rangle _{1}$ and $%
\left\vert f\right\rangle _{1}$, respectively. Also, by applying classical
pulses to the operation qubits ($2,3,...,n$), one can convert the states $%
\left\vert g\right\rangle $ and $\left\vert e\right\rangle $ of each of
these operation qubits into the states $\left\vert +\right\rangle $ and $%
\left\vert -\right\rangle $, respectively. Here and below, $\left\vert \pm
\right\rangle =\left( \left\vert g\right\rangle \pm \left\vert
e\right\rangle \right) /\sqrt{2}$ are two orthogonal states. Thus, the GHZ
state $\left\vert \text{GHZ}\right\rangle _{12...n}$ of the operation qubits
can be written as
\begin{equation}
\left\vert \text{GHZ}\right\rangle _{12...n}=\alpha
\prod_{l=2}^{n}\left\vert +\right\rangle _{l}\left\vert e\right\rangle
_{1}+\beta \prod_{l=2}^{n}\left\vert -\right\rangle _{l}\left\vert
f\right\rangle _{1}.
\end{equation}

Assume that the $2n$ memory qubits ($1^{\prime },2^{\prime },...,n^{\prime
},1^{\prime \prime },2^{\prime \prime },...,n^{\prime \prime }$) are in the
state $\left\vert e\right\rangle _{1^{\prime }}\prod_{l^{\prime }=2^{\prime
}}^{n^{\prime }}\left\vert +\right\rangle _{l^{\prime }}\prod_{l^{^{\prime
\prime }}=1^{\prime \prime }}^{n^{\prime \prime }}\left\vert -\right\rangle
_{l^{^{\prime \prime }}},$ which can be prepared by applying classical
pulses to the memory qubits ($1^{\prime },2^{\prime },...,n^{\prime
},1^{\prime \prime },2^{\prime \prime },...,n^{\prime \prime }$) each
initially in the ground state $\left\vert g\right\rangle $. Suppose that the
cavity is initially in a vacuum state $\left\vert 0\right\rangle _{c}$.
Thus, the initial state of the whole system is given by
\begin{equation}
\left( \alpha \prod_{l=2}^{n}\left\vert +\right\rangle _{l}\left\vert
e\right\rangle _{1}+\beta \prod_{l=2}^{n}\left\vert -\right\rangle
_{l}\left\vert f\right\rangle _{1}\right) \left\vert e\right\rangle
_{1^{\prime }}\prod_{l^{\prime }=2^{\prime }}^{n^{\prime }}\left\vert
+\right\rangle _{l^{\prime }}\prod_{l^{^{\prime \prime }}=1^{\prime \prime
}}^{n^{\prime \prime }}\left\vert -\right\rangle _{l^{^{\prime \prime
}}}\left\vert 0\right\rangle _{c}.
\end{equation}

\begin{figure}[tbp]
\begin{center}
\includegraphics[bb=49 188 621 436, width=10.5 cm, clip]{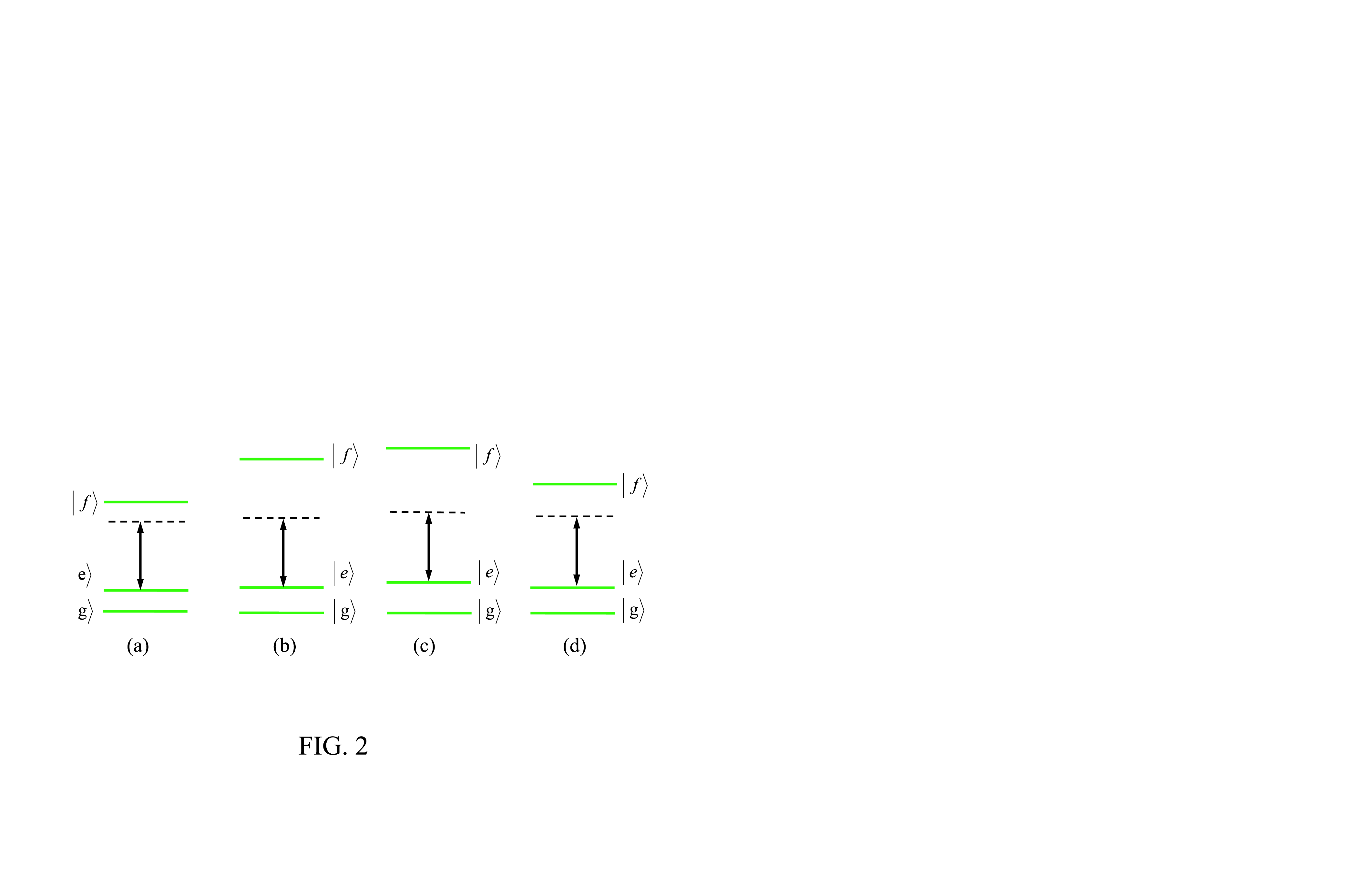} \vspace*{%
-0.08in}
\end{center}
\caption{(Color online) Level structure of each qubit before the GHZ state
transfer. (a), (b), (c), and (d) represent the level structures for qubit 1,
qubits ($2,3,...,n$), qubits ($2^{\prime },3^{\prime },...,n^{\prime
},1^{\prime \prime },2^{\prime \prime },...,n^{\prime \prime }$), and qubit $%
1^{\prime }$, respectively. Each double-arrow vertical line represents the
cavity frequency (not adjusted), which is highly detuned from the transition
frequency between any two levels of each qubit. Thus, each qubit is
initially decoupled from the cavity before the GHZ state transfer. The
difference between the level structures of (a) and (b) [(c) and (d)] can be
readily achieved by adjusting the level spacings of qubit 1 ( qubit $%
1^{\prime }$) before the GHZ state transfer [28,39,42-46]. For simplicity,
here and in Fig.~3, we consider the case that the spacing between the levels
$\left\vert g\right\rangle$ and $\left\vert e\right\rangle $ is smaller than
that between the levels $\left\vert e\right\rangle$ and $\left\vert
f\right\rangle $. Alternatively, the spacing between the levels $\left\vert
g\right\rangle$ and $\left\vert e\right\rangle $ can be larger than that
between the levels $\left\vert e\right\rangle$ and $\left\vert
f\right\rangle $.}
\label{fig:2}
\end{figure}

\begin{figure}[tbp]
\begin{center}
\includegraphics[bb=49 188 621 436, width=10.5 cm, clip]{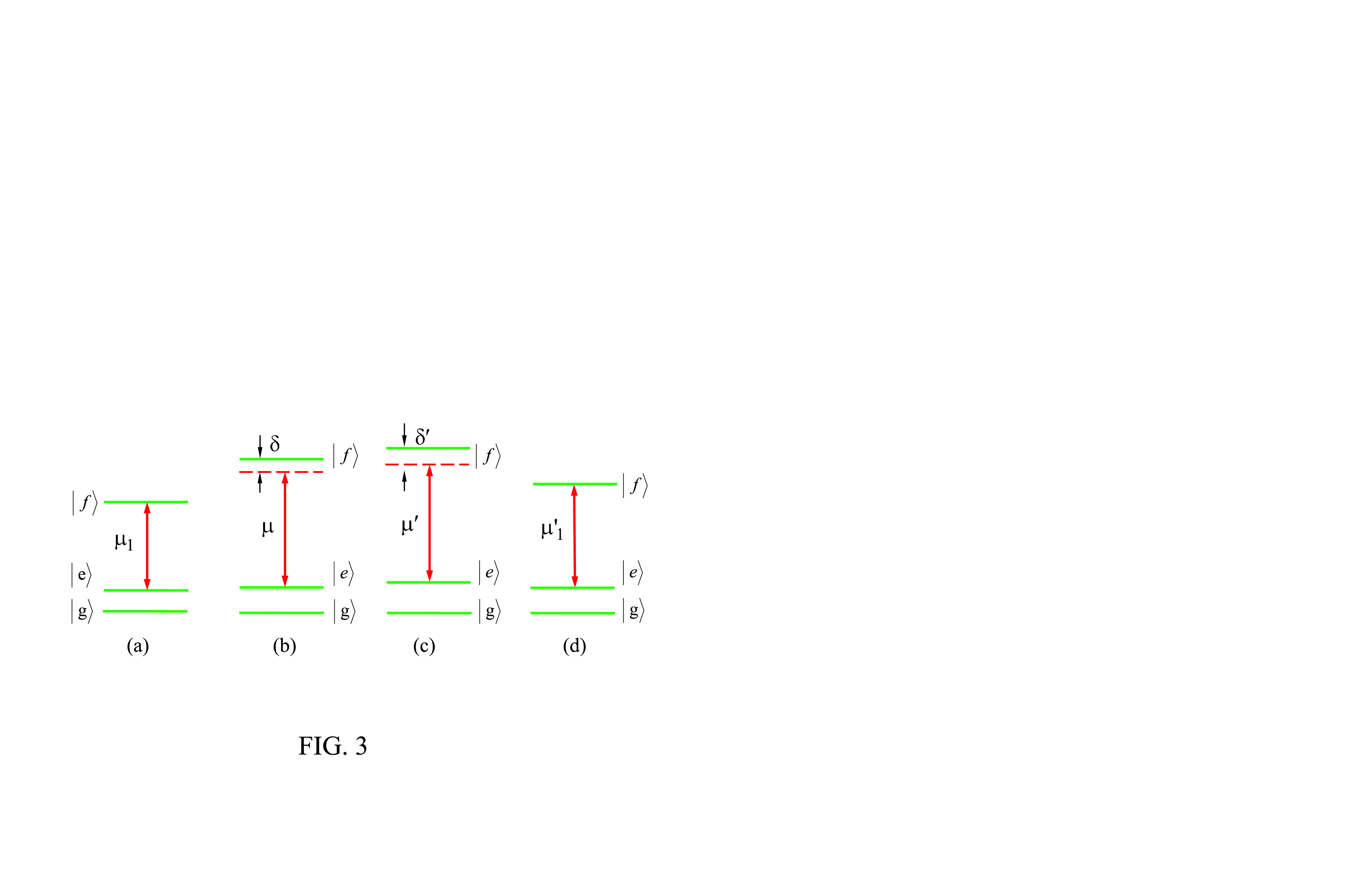} \vspace*{%
-0.08in}
\end{center}
\caption{(Color online) Illustration of qubit-cavity interaction. (a) The
cavity resonantly interacts with the $\left\vert e\right\rangle
\leftrightarrow \left\vert f\right\rangle $ transition of qubit 1. (b) The
cavity dispersively interacts with the $\left\vert e\right\rangle
\leftrightarrow \left\vert f\right\rangle $ transition of qubits ($2,3,...,n$%
). (c) The cavity dispersively interacts with the $\left\vert e\right\rangle
\leftrightarrow \left\vert f\right\rangle $ transition of qubits ($2^{\prime
},3^{\prime },...,n^{\prime },1^{\prime \prime},2^{\prime \prime
},...,n^{\prime \prime }$). (d) The cavity resonantly interacts with the $%
\left\vert e\right\rangle \leftrightarrow \left\vert f\right\rangle $
transition of qubit $1^{\prime }$. The level $\left\vert g\right\rangle$ of
each qubit is not affected by the cavity as long as the $\left\vert
g\right\rangle \leftrightarrow \left\vert e\right\rangle $ and $\left\vert
g\right\rangle \leftrightarrow \left\vert f\right\rangle $ transitions are
forbidden or highly detuned (i.e., decoupled) from the cavity mode}
\label{fig:3}
\end{figure}

For simplicity, in the following we will use the phrase \textquotedblleft $%
\left\vert e\right\rangle \leftrightarrow \left\vert f\right\rangle $
transition\textquotedblright\ to represent the \textquotedblleft transition
between the levels $\left\vert e\right\rangle $ and $\left\vert
f\right\rangle $\textquotedblright . Similar phrases are used to represent
the transition between the levels $\left\vert g\right\rangle $ and $%
\left\vert e\right\rangle $ as well as the transition between the levels $%
\left\vert g\right\rangle $ and $\left\vert f\right\rangle .$

From the description given below, one can see that there is no need of
adjusting the qubit level spacings during the entire GHZ state transfer. The
whole procedure for transferring the GHZ state $\left\vert \text{GHZ}%
\right\rangle _{12...n}$ of the $n$ operation qubits ($1,2,...,n$) onto the $%
2n$ memory qubits ($1^{\prime },2^{\prime },...,n^{\prime },1^{\prime \prime
},2^{\prime \prime },...,n^{\prime \prime }$) is listed as follows:

Step (i): Bring the cavity resonant with the $\left\vert e\right\rangle
\leftrightarrow \left\vert f\right\rangle $ transition of qubit $1$ [Fig.
3(a)]. The Hamiltonian describing this operation is given by $H_{I}=\hbar
\left( \mu _{1}a^{+}\left\vert e\right\rangle _{1}\left\langle f\right\vert
\right) +h.c.,$ where $a^{+}$ is the photon creation operator of the cavity
mode, and $\mu _{1}$ is the resonant coupling constant between the cavity
and the $\left\vert e\right\rangle \leftrightarrow \left\vert f\right\rangle
$ transition of qubit $1$. Under the Hamiltonian $H_{I}$, the state
component $\left\vert e\right\rangle _{1}\left\vert 0\right\rangle _{c}$ is
not changed because of $H_{I}\left\vert e\right\rangle _{1}\left\vert
0\right\rangle _{c}=0$. However, after an interaction time $t_{1}=\pi /(2\mu
_{1})$ (i.e., half a Rabi oscillation), the state $\left\vert f\right\rangle
_{1}\left\vert 0\right\rangle _{c}$ changes to $-i\left\vert e\right\rangle
_{1}\left\vert 1\right\rangle _{c}$ (for the details, see [47]). Hence, the
initial state (1) of the whole system becomes
\begin{equation}
\left( \alpha \prod_{l=2}^{n}\left\vert +\right\rangle _{l}\left\vert
0\right\rangle _{c}-i\beta \prod_{l=2}^{n}\left\vert -\right\rangle
_{l}\left\vert 1\right\rangle _{c}\right) \left\vert e\right\rangle
_{1}\left\vert e\right\rangle _{1^{\prime }}\prod_{l^{\prime }=2^{\prime
}}^{n^{\prime }}\left\vert +\right\rangle _{l^{\prime }}\prod_{l^{^{\prime
\prime }}=1^{\prime \prime }}^{n^{\prime \prime }}\left\vert -\right\rangle
_{l^{^{\prime \prime }}}.
\end{equation}
Step (ii): Bring the cavity dispersively coupled with the $\left\vert
e\right\rangle \leftrightarrow \left\vert f\right\rangle $ transition of
operation qubits ($2,3,...,n$), with non-resonant coupling strength $\mu $
and detuning $\delta $$=\omega _{fe}-\omega _{c}$ [Fig. 3(b)]. Meanwhile,
the cavity is dispersively coupled to the $\left\vert e\right\rangle
\leftrightarrow \left\vert f\right\rangle $ transition of memory qubits ($%
2^{\prime },3^{\prime },...,n^{\prime },1^{\prime \prime },2^{\prime \prime
},...,n^{\prime \prime }$), with non-resonant coupling strength $\mu
^{\prime }$ and detuning $\delta ^{\prime }$$=\omega _{fe}^{\prime }-\omega
_{c}$ [Fig.~3(c)]. Here, $\omega _{fe}$ ($\omega _{fe}^{\prime }$) is the
transition frequency between the levels $\left\vert e\right\rangle $ and $%
\left\vert f\right\rangle $ for operation qubits ($2,3,...,n$) [memory
qubits ($2^{\prime },3^{\prime },...,n^{\prime },1^{\prime \prime
},2^{\prime \prime },...,n^{\prime \prime })$], $\omega _{c}$ is the
(adjusted)\ cavity frequency. In the interaction picture, the Hamiltonian is
given by (setting $\hbar =1$)%
\begin{eqnarray}
H &=&\mu \sum\limits_{l=2}^{n}\left( e^{-i\delta t}\left\vert e\right\rangle
_{l}\left\langle f\right\vert a^{+}+h.c.\right) +\mu ^{\prime
}\sum\limits_{l^{\prime }=2^{\prime }}^{n^{\prime }}\left( e^{-i\delta
^{\prime }t}\left\vert e\right\rangle _{l^{\prime }}\left\langle
f\right\vert a^{+}+h.c.\right)  \notag \\
&&+\mu ^{\prime }\sum\limits_{l^{^{\prime \prime }}=1^{^{\prime \prime
}}}^{n^{\prime \prime }}\left( e^{-i\delta ^{\prime }t}\left\vert
e\right\rangle _{l^{^{\prime \prime }}}\left\langle f\right\vert
a^{+}+h.c.\right) .
\end{eqnarray}%
Under the large detuning conditions $\delta \gg \mu $ and $\delta ^{\prime
}\gg \mu ^{\prime },$ there is no energy exchange between the qubits and the
cavity mode. Then, the system dynamics described by the Hamiltonian of Eq.
(5) is equivalent to that determined by the following effective Hamiltonian
[48-50]
\begin{eqnarray}
H &=&\lambda \sum\limits_{l=2}^{n}\left( \left\vert f\right\rangle
_{l}\left\langle f\right\vert aa^{+}-\left\vert e\right\rangle
_{l}\left\langle e\right\vert a^{+}a\right)  \notag \\
&&+\lambda ^{\prime }\sum\limits_{l^{\prime }=2^{\prime }}^{n^{\prime
}}\left( \left\vert f\right\rangle _{l^{\prime }}\left\langle f\right\vert
aa^{+}-\left\vert e\right\rangle _{l^{\prime }}\left\langle e\right\vert
a^{+}a\right)  \notag \\
&&+\lambda ^{\prime }\sum\limits_{l^{^{\prime \prime }}=1^{\prime \prime
}}^{n^{\prime \prime }}\left( \left\vert f\right\rangle _{l^{^{\prime \prime
}}}\left\langle f\right\vert aa^{+}-\left\vert e\right\rangle _{l^{^{\prime
\prime }}}\left\langle e\right\vert a^{+}a\right)  \notag \\
&&+\lambda \sum_{l,k=2}^{n}\left\vert f\right\rangle _{l}\left\langle
e\right\vert \otimes \left\vert e\right\rangle _{k}\left\langle f\right\vert
\notag \\
&&+\lambda ^{\prime }\sum_{l^{\prime },k^{\prime }=2^{\prime }}^{n^{\prime
}}\left\vert f\right\rangle _{l^{\prime }}\left\langle e\right\vert \otimes
\left\vert e\right\rangle _{k^{\prime }}\left\langle f\right\vert  \notag \\
&&+\lambda ^{\prime }\sum_{l^{^{\prime \prime }},k^{^{\prime \prime
}}=1^{\prime \prime }}^{n^{^{\prime \prime }}}\left\vert f\right\rangle
_{l^{^{\prime \prime }}}\left\langle e\right\vert \otimes \left\vert
e\right\rangle _{k^{^{\prime \prime }}}\left\langle f\right\vert  \notag \\
&&+\widetilde{\lambda }\sum_{l=2,k^{\prime }=2^{\prime }}^{n,n^{\prime
}}e^{i\left( \delta -\delta ^{\prime }\right) t}\left\vert f\right\rangle
_{l}\left\langle e\right\vert \otimes \left\vert e\right\rangle _{k^{\prime
}}\left\langle f\right\vert  \notag \\
&&+\widetilde{\lambda }\sum_{l=2,k^{^{\prime \prime }}=1^{^{\prime \prime
}}}^{n,n^{^{\prime \prime }}}e^{i\left( \delta -\delta ^{\prime }\right)
t}\left\vert f\right\rangle _{l}\left\langle e\right\vert \otimes \left\vert
e\right\rangle _{k^{^{\prime \prime }}}\left\langle f\right\vert  \notag \\
&&+\lambda ^{\prime }\sum_{l^{^{\prime }}=2^{\prime },k^{^{\prime \prime
}}=1^{\prime \prime }}^{n^{\prime },n^{^{\prime \prime }}}\left\vert
f\right\rangle _{l^{^{\prime }}}\left\langle e\right\vert \otimes \left\vert
e\right\rangle _{k^{^{\prime \prime }}}\left\langle f\right\vert ,
\end{eqnarray}%
where $\lambda =\mu ^{2}/\delta ,$ $\lambda ^{\prime }=\left( \mu ^{\prime
}\right) ^{2}/\delta ^{\prime },$ $\widetilde{\lambda }=\frac{\delta \delta
^{\prime }}{2}\left( \frac{1}{\delta }+\frac{1}{\delta ^{\prime }}\right) ,$
$\quad l\neq k$ (line 4), $l^{\prime }\neq k^{\prime }$ (line 5), and $%
l^{\prime \prime }\neq k^{\prime \prime }$ (line 6). The terms in lines 1,
2, and 3 of Eq.~(6) describe the photon-number dependent Stark shifts. The
term in line 4 describes the \textquotedblleft dipole\textquotedblright\
couplings between the $l$th operation qubit and the $k$th operation qubit.
The term in line 5 represents the \textquotedblleft
dipole\textquotedblright\ couplings between the $l^{\prime }$th memory qubit
and the $k^{\prime }$th memory qubit. The term in line 6 denotes the
\textquotedblleft dipole\textquotedblright\ couplings between the $l^{\prime
\prime }$th memory qubit and the $k^{\prime \prime }$th memory qubit. The
terms in the last three lines describe the \textquotedblleft
dipole\textquotedblright\ coupling between the $l$th operation qubit and the
$k^{\prime }$th memory qubit, the $l$th operation qubit and the $k^{\prime
\prime }$th memory qubit, as well as the $l^{\prime }$th memory qubit and
the $k^{\prime \prime }$th memory qubit, mediated by the cavity. Note that
the level $\left\vert f\right\rangle $ of each qubit is not involved in Eq.
(4). Thus, one can easily find that only the terms $-\lambda
\sum\limits_{l=2}^{n}\left\vert e\right\rangle _{l}\left\langle e\right\vert
a^{+}a,$ $-\lambda ^{\prime }\sum\limits_{l^{\prime }=2^{\prime
}}^{n^{\prime }}\left\vert e\right\rangle _{l^{\prime }}\left\langle
e\right\vert a^{+}a,$ and $-\lambda ^{\prime }\sum\limits_{l^{\prime \prime
}=1^{\prime \prime }}^{n^{\prime \prime }}\left\vert e\right\rangle
_{l^{^{\prime \prime }}}\left\langle e\right\vert a^{+}a$ of Eq. (6) have
contribution to the time evolution of the state (4), while all other terms
in Eq. (6) acting on the state (4) result in zero. In other words, with
respective to the state (4), the Hamiltonian~(6) reduces to
\begin{equation}
H=-\lambda \sum\limits_{l=2}^{n}\left\vert e\right\rangle _{l}\left\langle
e\right\vert a^{+}a-\lambda ^{\prime }\sum\limits_{l=2^{\prime }}^{n^{\prime
}}\left\vert e\right\rangle _{l^{\prime }}\left\langle e\right\vert
a^{+}a-\lambda ^{\prime }\sum\limits_{l^{\prime \prime }=1^{\prime \prime
}}^{n^{\prime \prime }}\left\vert e\right\rangle _{l^{^{\prime \prime
}}}\left\langle e\right\vert a^{+}a.
\end{equation}%
Under the Hamiltonian (7), the state (4)\ evolves into%
\begin{eqnarray}
&&\left[ \alpha \prod_{l=2}^{n}\left\vert +\right\rangle
_{l}\prod_{l^{\prime }=2^{\prime }}^{n^{\prime }}\left\vert +\right\rangle
_{l^{\prime }}\prod_{l^{^{\prime \prime }}=1^{\prime \prime }}^{n^{\prime
\prime }}\left\vert -\right\rangle _{l^{^{\prime \prime }}}\left\vert
0\right\rangle _{c}-i\beta \prod_{l=2}^{n}\left( \left\vert g\right\rangle
_{l}-e^{i\lambda t}\left\vert e\right\rangle _{l}\right) \right.  \notag \\
&&\left. \times \prod_{l^{\prime }=2^{\prime }}^{n^{\prime }}\left(
\left\vert g\right\rangle _{l^{\prime }}+e^{i\lambda ^{\prime }t}\left\vert
e\right\rangle _{l^{\prime }}\right) \prod_{l^{\prime \prime }=1^{\prime
\prime }}^{n^{\prime \prime }}\left( \left\vert g\right\rangle _{l^{\prime
\prime }}-e^{i\lambda ^{\prime }t}\left\vert e\right\rangle _{l^{\prime
\prime }}\right) \left\vert 1\right\rangle _{c}\right] \left\vert
e\right\rangle _{1}\left\vert e\right\rangle _{1^{\prime }}
\end{eqnarray}%
In the case of $t_{2}=\left( 2m+1\right) \pi /\lambda =\left( 2k+1\right)
\pi /\lambda ^{\prime }$ ($m$ and $k$ are zero or positive integers)$,$ we
have from Eq.~(8)
\begin{equation}
\left[ \alpha \prod_{l^{\prime }=2^{\prime }}^{n^{\prime }}\left\vert
+\right\rangle _{l^{\prime }}\prod_{l^{^{\prime \prime }}=1^{\prime \prime
}}^{n^{\prime \prime }}\left\vert -\right\rangle _{l^{^{\prime \prime
}}}\left\vert 0\right\rangle _{c}-i\beta \prod_{l^{\prime }=2^{\prime
}}^{n^{\prime }}\left\vert -\right\rangle _{l^{\prime }}\prod_{l^{^{\prime
\prime }}=1^{\prime \prime }}^{n^{\prime \prime }}\left\vert +\right\rangle
_{l^{^{\prime \prime }}}\left\vert 1\right\rangle _{c}\right]
\prod_{l=2}^{n}\left\vert +\right\rangle _{l}\left\vert e\right\rangle
_{1}\left\vert e\right\rangle _{1^{\prime }}.
\end{equation}
Step (iii): Bring the cavity resonant with the $\left\vert e\right\rangle
\leftrightarrow \left\vert f\right\rangle $ transition of qubit $1^{\prime }$
[Fig.~3(d)]. The Hamiltonian describing this operation is given by $%
H_{I}=\hbar \left( \mu _{1^{\prime }}a^{+}\left\vert e\right\rangle
_{1^{\prime }}\left\langle f\right\vert \right) +h.c.,$ where $\mu
_{1^{\prime }}$ is the resonant coupling constant between the resonator and
the $\left\vert e\right\rangle \leftrightarrow \left\vert f\right\rangle $
transition of qubit $1^{\prime }$. After an interaction time $t_{3}=3\pi
/(2\mu _{1^{\prime }}),$ the resonator mode and qubit $1^{\prime }$ undergo
the transformation $\left\vert e\right\rangle _{1^{\prime }}\left\vert
0\right\rangle _{c}\rightarrow \left\vert e\right\rangle _{1^{\prime
}}\left\vert 0\right\rangle _{c}$ and $\left\vert e\right\rangle _{1^{\prime
}}\left\vert 1\right\rangle _{c}\rightarrow i\left\vert f\right\rangle
_{1^{\prime }}\left\vert 0\right\rangle _{c}$. Thus, the state (9) becomes
\begin{equation}
\left[ \alpha \prod_{l^{\prime }=2^{\prime }}^{n^{\prime }}\left\vert
+\right\rangle _{l^{\prime }}\prod_{l^{^{\prime \prime }}=1^{\prime \prime
}}^{n^{\prime \prime }}\left\vert -\right\rangle _{l^{^{\prime \prime
}}}\left\vert e\right\rangle _{1^{\prime }}+\beta \prod_{l^{\prime
}=2^{\prime }}^{n^{\prime }}\left\vert -\right\rangle _{l^{\prime
}}\prod_{l^{^{\prime \prime }}=1^{\prime \prime }}^{n^{\prime \prime
}}\left\vert +\right\rangle _{l^{^{\prime \prime }}}\left\vert
f\right\rangle _{1^{\prime }}\right] \prod_{l=2}^{n}\left\vert
+\right\rangle _{l}\left\vert e\right\rangle _{1}\left\vert 0\right\rangle
_{c}.
\end{equation}%
By comparing Eq.~(10) with Eq.~(3) or Eqs.~(1-2), one can see that the
original $n$-qubit GHZ state of qubits ($1,2,...,n$) has been transferred
onto the memory qubits ($1^{\prime },2^{\prime },...,n^{\prime },1^{\prime
\prime },2^{\prime \prime },...,n^{\prime \prime }$) after the operation.
After this step of operation, the cavity frequency needs to be adjusted back
to its original frequency (Fig. 2) such that the qubit system is decoupled
from the cavity.

Note that by applying classical pulses to the qubit $1^{\prime },$ the
states $\left\vert e\right\rangle _{1^{\prime }}$ and $\left\vert
f\right\rangle _{1^{\prime }}$ can be easily converted into the states $%
\left\vert g\right\rangle _{1^{\prime }}$ and $\left\vert e\right\rangle
_{1^{\prime }}$, respectively. Also, by applying classical pulses to the
memory qubits ($2^{\prime },...,n^{\prime },1^{\prime \prime },2^{\prime
\prime },...,n^{\prime \prime }$), one can convert the states $\left\vert
+\right\rangle $ and $\left\vert -\right\rangle $ of each of the memory
qubits ($2^{\prime },...,n^{\prime },1^{\prime \prime },2^{\prime \prime
},...,n^{\prime \prime }$) into the states $\left\vert g\right\rangle $ and $%
\left\vert e\right\rangle $, respectively. In this case, the entangled GHZ
state of the $2n$ memory qubits, i.e., the left part in the square bracket of
Eq.~(10), will become
\begin{eqnarray}
\left\vert \text{GHZ}\right\rangle _{DF} &=&\alpha \prod_{l^{\prime
}=1^{\prime }}^{n^{\prime }}\left\vert g\right\rangle _{l^{\prime
}}\prod_{l^{^{\prime \prime }}=1^{\prime \prime }}^{n^{\prime \prime
}}\left\vert e\right\rangle _{l^{^{\prime \prime }}}+\beta \prod_{l^{\prime
}=1^{\prime }}^{n^{\prime }}\left\vert e\right\rangle _{l^{\prime
}}\prod_{l^{^{\prime \prime }}=1^{\prime \prime }}^{n^{\prime \prime
}}\left\vert g\right\rangle _{l^{^{\prime \prime }}}  \notag \\
&=&\alpha \left\vert ge\right\rangle _{1^{\prime }1^{\prime \prime
}}\left\vert ge\right\rangle _{2^{\prime }2^{\prime \prime }}...\left\vert
ge\right\rangle _{n^{\prime }n^{\prime \prime }}+\beta \left\vert
eg\right\rangle _{1^{\prime }1^{\prime \prime }}\left\vert eg\right\rangle
_{2^{\prime }2^{\prime \prime }}...\left\vert eg\right\rangle _{n^{\prime
}n^{\prime \prime }}.
\end{eqnarray}%
The state (11) is exactly a GHZ entangled state encoded within a DFS against
collective dephasing, which is spanned by the basis vectors $\left\vert
ge\right\rangle $ and $\left\vert eg\right\rangle $ of every two of paired
memory qubits \{$1^{\prime },1^{\prime \prime }$\}, \{$2^{\prime },2^{\prime
\prime }$\},......, \{$n^{\prime },n^{\prime \prime }$\}. When the two
memory qubits in each pair undergo a collective decoherence, this encoded
GHZ state (11) is immune to a phase-dumping environment because of $%
H\left\vert \text{GHZ}\right\rangle _{DF}=0$, where $H=\sum_{j=1}^{n}g_{j}E%
\otimes \left( \sigma _{j^{\prime },z}+\sigma _{j^{\prime \prime },z}\right)
$ is an interaction Hamiltonian describing the $2n$ memory qubits
interacting with a dephasing environment [31-33,37]. Here, $E$ is an
arbitrary environment operator, $g_{j}$ is the coupling constant of the two
paired memory qubits ($j^{\prime },j^{\prime \prime }$) with the noise
environment, $\sigma _{j^{\prime },z}=\left\vert e\right\rangle
_{j^{\prime }}\left\langle e\right\vert $ $-\left\vert g\right\rangle
_{j^{\prime }}\left\langle g\right\vert $ and $\sigma _{j^{\prime \prime
},z}=\left\vert e\right\rangle _{j^{\prime \prime }}\left\langle
e\right\vert $ $-\left\vert g\right\rangle _{j^{\prime \prime }}\left\langle
g\right\vert $ are the operators of qubits $j^{\prime }$ and $j^{\prime
\prime }$ respectively. By comparing Eq.(1) with Eq.(11), one can see that
only two memory qubits are needed to protect one of GHZ qubits against
decoherence caused by the dephasing environment.

Tuning the cavity frequency has been reported in various experiments for
both optical cavities and microwave cavities [51-54]. For instance, the
rapid tuning of cavity frequencies in $1-3$ nanoseconds has been
experimentally demonstrated for a superconducting transmission line
resonator [53,54]. We should mention that adjusting the cavity frequency is
unnecessary for this proposal. Alternatively, the qubit-cavity resonant
interaction and the qubit-cavity dispersive interaction, which are required
by the GHZ state transfer, can be achieved by adjusting the level spacings
of the qubits [28,39,42-46].

Several points may need to be addressed here. First, the GHZ state transfer
only employs the coupling of the cavity with the $\left\vert e\right\rangle
\leftrightarrow $ $\left\vert f\right\rangle $ transition of each qubit
(Fig.~3). Hence, this proposal is applicable to qubits with forbidden $%
\left\vert g\right\rangle \leftrightarrow \left\vert e\right\rangle $ and $%
\left\vert g\right\rangle \leftrightarrow \left\vert f\right\rangle $
transitions. It also applies to qubits with $\left\vert g\right\rangle
\leftrightarrow \left\vert e\right\rangle $ and $\left\vert g\right\rangle
\leftrightarrow \left\vert f\right\rangle $ transitions as long as these
transitions are highly detuned (i.e., decoupled) from the cavity, which can
be achieved by adjusting the qubits' level spacings before performing the
GHZ state transfer [28,39,42-46]. Second, as shown above, the level $%
\left\vert f\right\rangle $ for operation qubits ($2,3,...,n$) and memory
qubits ($2^{\prime },3^{\prime },...,n^{\prime },1^{\prime \prime
},2^{\prime \prime },...,n^{\prime \prime }$) is unpopulated, i.e., the
level $\left\vert f\right\rangle $ is occupied only for the two qubits $1$
and $1^{\prime }$ during the entire operation. Third, the operations for
transferring GHZ states have nothing to do with $\alpha $ and $\beta .$
Last, the method is applicable to a 1D, 2D or 3D cavity or resonator as long
as the conditions described above can be met.

We should mention that since no measurement is involved during the GHZ state
transfer, it is straightforward to show that the transferred GHZ state can
be transferred back onto the operation qubits from the memory qubits by
performing the inverse operations of the above unitary operations.

Note that the GHZ state given in Eq. (1) is an $n$-qubit quantum
secret sharing code for encoding an arbitrary single-qubit pure state $\alpha \left\vert
g\right\rangle +\beta \left\vert e\right\rangle $ [55], while the transferred GHZ state
given in Eq.~(11) is a quantum secret sharing encoded with $n$ pairs of memory qubits
in a DFS. Thus, this proposal also provides a way to transfer quantum secret sharing
from $n$ operation qubits to $n$ pairs of memory qubits in a DFS. The state
in Eq.~(11) may also be usable for keeping a resource of controlled quantum teleportation.

Note that a SWAP gate on qubits $a$ and $b$, described by the state
transformation $\left\vert i\right\rangle _{a}\left\vert j\right\rangle
_{b}\rightarrow \left\vert j\right\rangle _{a}\left\vert i\right\rangle _{b}$
with $i,j\in \{0,1\},$ can be used to implement the state transfer from
qubit $a$ to qubit $b$ or vice versa. Thus, it is straightforward to show that the transfer of
an $n$-qubit GHZ state, considered in this work, can be completed with $n$
SWAP gates each acting on an operation qubit and a memory \textquotedblleft
logical qubit\textquotedblright . Here, a memory logical qubit consists of
two memory qubits (e.g., qubit pair $1^{\prime }1^{\prime \prime },2^{\prime
}2^{\prime \prime },...,$ or $n^{\prime }n^{\prime \prime }$), whose two
logical states are the states $\left\vert ge\right\rangle $ and $\left\vert
eg\right\rangle $ which form a DFS against collective dephasing. According
to [56], construction of a SWAP gate requires three CNOT gates, which can be
realized by the idea presented in [57]. In this sense, 3n CNOT
gates are needed for constructing $n$ SWAP gates. Thus, at least 3n basic
operations are necessary to transfer an $n$-qubit GHZ state from the
$n$ operation qubits to the $2n$ memory qubits, provided that a CNOT gate can be
realized with one basic operation only. In stark contrast, as shown above,
the present proposal requires only a few operational steps to transfer an $n$%
-qubit GHZ state, which is independent of $n$.

Based on cavity/circuit QED, there exist a number of schemes for generating the GHZ states
in various types of qubits (e.g., Refs. [12-18]), which are based on qubit-cavity resonant interaction (QCRI) or
dispersive interaction. As shown above, the GHZ state transfer is mainly based on step (ii),
which was performed via the dispersive interaction. Regarding the operation speed,
the GHZ state transfer is comparable to the GHZ state preparation based on
dispersive interaction. This is because the GHZ state generation requires an initial state preparation
of the qubit system.

When the number of qubits is large, the operational speed for the GHZ state transfer here
may also be comparable to that of the GHZ state generation based on QCRI. Reasons for that are the following.
The GHZ state preparation via QCRI requires a cavity photon resonantly interacting with each qubit sequentially,
and thus step-by-step control is needed. Although each step can be performed fast due to using QCRI,
the total operation time increases linearly with the number of qubits and becomes long
when the number of qubits is large. However, as shown above, the procedure for the GHZ state transfer only needs a
few operational steps and the operation time is independent of the number of qubits.

This work is also of interest from the following point of view. As mentioned in the introduction,
we considered two sets of qubits, i.e., operation qubits and memory qubits. The operation qubits are readily
controlled and thus preferable to be used for performing various quantum operations required by quantum computing.
Compared to the memory qubits, the operation qubits have shorter decoherence time. Thus, the prepared GHZ states with the operation qubits
are easy to be destroyed due to decoherence. One would need to repeat preparing them once
they are destroyed. However, this problem is mitigated by transferring the GHZ states from the operation qubits onto the memory qubits
for a long time storage.

\section{Discussion}

For the method to work, the following requirements need to be satisfied:

(i) The condition $\left( 2m+1\right) \pi /\lambda =(2k+1)\pi /\lambda
^{\prime }$ needs to be met. Because of $\lambda =\mu ^{2}/\delta $ and $%
\lambda ^{\prime }=\left( \mu ^{\prime }\right) ^{2}/\delta ^{\prime },$
this condition can be readily reached with an appropriate choice of $\delta $
(or $\delta ^{\prime }$) via adjusting the level spacings of qubits ($%
2,3,...,n$) [or qubits ($2^{\prime },3^{\prime },...,n^{\prime },1^{\prime
\prime },2^{\prime \prime },...,n^{\prime \prime }$)] or the coupling
constant $\mu $ ($\mu ^{\prime }$) by varying the qubit locations in the
cavity.

(ii) The total operation time is
\begin{equation}
\tau =\frac{\pi }{2\mu _{1}}+\frac{3\pi }{2\mu _{1^{\prime }}}+\frac{\left(
2m+1\right) \pi }{\lambda }+\tau _{p}+4\tau _{d},
\end{equation}%
which is independent of the number of qubits. Here, $t_{d}$ is the typical
time required for adjusting the cavity frequency while $\tau _{p}$ is the
typical time required for applying the classical pulses. The operation time $%
\tau $ should be much smaller than the energy relaxation time and the
dephasing time of qubits. In addition, $\tau $ should be much smaller than
the lifetime of the cavity mode, which is given by $\kappa ^{-1}=Q/\omega
_{c}$ ($Q$ is the quality factor of the cavity). In principle, these
requirements can be satisfied as follows. The operation time $\tau $ can be
reduced by increasing the coupling constants $\mu $ and $\mu ^{\prime },$ by
shortening the pulse duration via increasing the pulse Rabi frequency, and
by rapidly adjusting the cavity frequency (e.g., $\tau _{d}$ $\sim $ $1\--3$
ns is the typical time for adjusting the frequency of a microwave cavity in
experiments [53,54]). In addition, $\kappa ^{-1}$ can be increased by
employing a high-$Q$ resonator. Note that the symbol \textquotedblleft $\sim
$\textquotedblright\ used here and below means \textquotedblleft
approximately equals to\textquotedblright .

(iii) During step (ii), the occupation probability $p$ of the level $%
\left\vert f\right\rangle $ for each of operation qubits ($2,3,...,n$) and
the occupation probability $p^{\prime }$ of the level $\left\vert
f\right\rangle $ for each of memory qubits ($2^{\prime },3^{\prime
},...,n^{\prime },1^{\prime \prime },2^{\prime \prime },...,n^{\prime \prime
}$) are given by [58]
\begin{eqnarray}
p &\simeq &\frac{4\mu ^{2}}{4\mu ^{2}+\delta ^{2}},\text{ }  \notag \\
p^{\prime } &\simeq &\frac{4\left( \mu ^{\prime }\right) ^{2}}{4\left( \mu
^{\prime }\right) ^{2}+(\delta ^{\prime })^{2}}.
\end{eqnarray}%
The occupation probabilities $p$ and $p^{\prime }$ need to be negligibly
small in order to reduce the operation error. For the choice of $\delta
=10\mu $ and $\delta ^{\prime }=10\mu ^{\prime },$ we have $p,$ $p^{\prime
}\sim 0.04$, which can be further reduced by increasing the ratio of $\delta
/\mu $ and $\delta ^{\prime }/\mu ^{\prime }.$

\begin{figure}[tbp]
\begin{center}
\includegraphics[bb=0 0 1187 621, width=8.5 cm, clip]{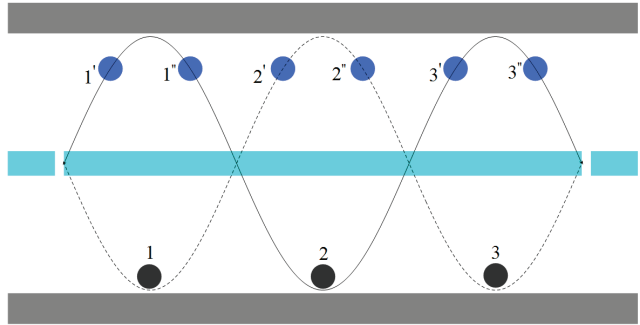} \vspace*{%
-0.08in}
\end{center}
\caption{(Color online) Setup for three superconducting flux qubits (black
dots), six superconducting transmon qubits (blue dots), and a
one-dimensional coplanar waveguide resonator. The two curved lines represent
the standing wave magnetic field of the resonator. Here, the flux qubits ($%
1,2,3$) act as the operation qubits , while the paired transmon qubits,
i.e., qubits \{$1^{\prime },1^{\prime \prime }$\}, qubits \{$2^{\prime
},2^{\prime \prime }$\}, and qubits \{$3^{\prime },3^{\prime \prime }$\},
play a role of the memory qubits.}
\end{figure}

For the sake of definitiveness, let us consider the experimental possibility
of transferring a three-qubit GHZ state from three identical superconducting
flux qubits to six identical superconducting transmon qubits embedded in a
one-dimensional transmission line resonator (TLR) (Fig. 4). Flux qubits have
stronger level anharmonicity (compared to transmon qubits) while transmon
qubits have relatively long decoherence time reported in experiments. Thus,
as an example, we here consider using flux qubits as operation qubits while
transmon qubits as memory qubits. As a rough estimate, assume $\mu _{1}\sim
\mu \sim \mu ^{\prime }\sim \mu _{1^{\prime }}\sim 2\pi \times 10$ MHz and $%
\delta \sim 10\mu ,$ $\delta ^{\prime }\sim 10\mu ^{\prime }$, resulting in $%
\lambda =\lambda ^{\prime }$ and thus $m=k=0$ chosen for the shortest
operation time $\tau $. In addition, assume $\tau _{d}\sim 2$ ns, and $\tau
_{p}\sim 10$ ns (readily available because a microwave pulse Rabi frequency $%
\Omega /2\pi \sim 300$ MHz has been reported in experiments [59]). Note that
the coupling strengths with the values chosen here are readily available in
experiments because a coupling strength $\sim 636$ MHz has been reported for
a flux qubit coupled to a TLR [60] and a coupling constant $\sim 220$ MHz
has been experimentally demonstrated for a transmon qubit coupled to a TLR
[61]. For the parameters chosen here, we have $\tau \sim 0.6$ $\mu $s, which
is much shorter than the early-reported decoherence time $6-20$ $\mu $s of the flux qubits
[62] and lifetime $\sim 50-80$ $\mu $s for the
transmon qubits [63,64]. For both flux qubits and transmon qubits, the
typical transition frequency between two neighbor levels $\left\vert
e\right\rangle $ and $\left\vert f\right\rangle $ can be made to be $5-10$
GHz. Thus, as an example, choose $\omega _{c}\sim 5$ GHz. In addition,
consider $Q\sim 5\times 10^{5},$ and thus we have $\kappa ^{-1}\sim 16$ $\mu
$s, which is much longer than the operation time $\tau \sim 0.6$ $\mu $s
given above. The required resonator quality factor here is achievable in
experiment because TLRs with a (loaded)\ quality factor $Q\sim 10^{6}$ have
been experimentally demonstrated [65,66]. The result presented here shows
that transferring a three-qubit GHZ state from superconducting flux qubits
onto superconducting transmon qubits in a DFS is possible within present-day
circuit QED. We remark that further investigation is needed for each
particular experimental setup. However, this requires a rather lengthy and
complex analysis, which is beyond the scope of this theoretical work.

\section{Conclusion}

We have shown that multi-qubit GHZ states can be deterministically transferred from the operation qubits
onto the memory qubits, through only a few basic operations. Since the transferred GHZ
states are encoded with memory qubits within a DFS, they are immune from dephasing environments,
and thus can in principle be protected against decoherence caused by dephasing for an
indefinite length of time (without the need for frequent checking). As shown
above, the state transfer does not depend on the number of qubits, the
operation time does not increase with the number of qubits, and no
measurement is needed during the entire operation. Moreover, since the level
$\left\vert f\right\rangle $ only for two qubits was occupied during the
operation, decoherence from the qubits is greatly suppressed. This proposal
can be applied to a wide range of hybrid qubits such as natural atoms and artificial atoms (e.g., quantum dots, NV centers, and
various superconducting qubits).

\section{Acknowledgments}

This work was supported in part by the Ministry of Science and Technology
of China under Grant No. 2016YFA0301802, the National Natural Science Foundation
of China under Grant Nos. 11074062 and 11374083, and the Zhejiang Natural
Science Foundation under Grant Nos. LZ13A040002 and LY15A040006. This work was also supported
by the funds of Hangzhou City for supporting the Hangzhou-City Quantum Information
and Quantum Optics Innovation Research Team.

\end{document}